\documentclass[
preprint,  % 練習用
superscriptaddress,
showpacs,
showkeys,
preprintnumbers,
 amsmath,amssymb,
 aps,
 pre,
]{revtex4-1}

\usepackage{graphicx}% Include figure files
\usepackage{dcolumn}% Align table columns on decimal point
\usepackage{bm}% bold math
\usepackage{hyperref}% add hypertext capabilities
\usepackage{enumerate}
\usepackage[utf8]{inputenc} 
\usepackage[mathlines]{lineno}% Enable numbering of text and display math
\newcommand{\affA}{
Department of Physics, The University of Tokyo, Komaba, Meguro, Tokyo 153-8505 
}

\renewcommand{\vec}{\bm}

\usepackage{varioref}

\begin{document}

\title{Tensor-network algorithm for nonequilibrium relaxation in the thermodynamic limit}

\author{Yoshihito Hotta}
\affiliation{\affA}
% \affiliation{\affB}
% author{Naomichi Hatano}
%{\textyen}affiliation{{\textyen}affB}
\email{hotta-y@mail.iis.u-tokyo.ac.jp}

%{\textyen}collaboration{MUSO Collaboration}%{\textyen}noaffiliation

\date{\today}% It is always {\textyen}today, today, but any date may be explicitly specified

\begin{abstract}
We propose a tensor-network algorithm for discrete-time stochastic dynamics of a homogeneous system in the thermodynamic limit.
We map a $d$-dimensional nonequilibrium Markov process to a $(d+1)$-dimensional infinite tensor network by using a higher-order singular-value decomposition.
%As applications of the algorithm, we analyze nonequilibrium relaxation of the one- and two-dimensional Ising models from ordered states to equilibrium and estimate the dynamical critical exponent $z=2.16(5)$ for the two-dimensional Ising model.
As an application of the algorithm, we compute the nonequilibrium relaxation from a fully magnetized state to equilibrium of the one- and two- dimensional Ising models with periodic boundary conditions. 
Utilizing the translational invariance of the systems, we analyze the behavior in the thermodynamic limit directly.
We estimated the dynamical critical exponent $z=2.16(5)$ for the two-dimensional Ising model.
Our approach fits well with the framework of the nonequilibrium-relaxation method.
Our algorithm can compute time evolution of the magnetization of a large system precisely for a relatively short period.
In the nonequilibrium-relaxation method, on the other hand, one needs to simulate dynamics of a large system for a short time.
The combination of the two provides a new approach to the study of critical phenomena.

% The nonequilibrium-relaxation method combined with our algorithm offers an alternative in a study of critical phenomena with tensor networks other than tensor renormalization group algorithms, which calculate order parameters at equilibrium states.

% We propose a tensor-network algorithm for simulation of Markov chain (discrete-time Markov process).
% As a concrete example, we consider kinetic Ising model throughout this paper.
% We map $d$-dimensional classical nonequilibrium system to $(d+1)$-dimensional equilibrium system 
% by representing a transition rate as a tensor-network operator by higher-order singular-value decomposition (HOSVD).
% Our method offers an alternative other than Monte Carlo method in simulating Markov chain. 
% This method will be particularly useful in computing a statistical variable with a large variance. 
% We calculate $\exp(-\beta W(t))$, which appears in Jarzynski's equality, as an example of a high-variance variable.
%{\textyen}begin{description}
%{\textyen}item[Usage]
%{\textyen}item[PACS numbers]
%May be entered using the {\textyen}verb+{\textyen}pacs{#1}+ command.
%{\textyen}item[Structure]
%You may use the {\textyen}texttt{description} environment to structure your abstract;
%use the optional argument of the {\textyen}verb+{\textyen}item+ command to give the category of each item. 
%{\textyen}end{description}
\end{abstract}

% {\textyen}pacs{{\textyen}begin{warning}Valid PACS appear here{\textyen}end{warning}}
%{\textyen}keywords{self-avoiding walk}
\maketitle

\section{Introduction}
It is important to develop algorithms to analyze stochastic processes using tensor networks. 
Stochastic processes often appear in statistical mechanics.
Monte Carlo methods are most often used to simulate stochastic processes. Density-matrix renormalization group (DMRG) is also sometimes used to analyze them~\cite{hieida98, carlon99, carlon01, nagy02, kemper01, temme10}.
We develop an algorithm to analyze stochastic processes using tensor networks in this paper. 
Tensor-network algorithms are  generalization of DMRG and transfer-matrix methods to higher dimensions~\cite{orus14, schollwoeck11, verstraete08} and can handle models in two and higher dimensions straightforwardly.
We combine our algorithm with the nonequilibrium-relaxation method~\cite{ito93,ozeki07} to evaluate critical exponents.
The computational time of our algorithm does not depend on the system size when the system is homogeneous, 
whereas the computational time of Monte Carlo simulation generally depends on the system size.

Monte Carlo methods are stochastic processes that are often used in studies of statistical mechanics.
Although Monte Carlo methods have advantages, such as wide applicability and easiness of implementation, they also have drawbacks, \textit{e.g.}, the dependence of computational complexity on the system size. 
As another drawback, equilibrium Monte Carlo analysis of critical phenomena become extremely difficult as the system approaches a critical point because of the divergence of the relaxation time. 
The nonequilibrium-relaxation method~\cite{ito93,ozeki07}, on the other hand, determines critical exponents including dynamical ones by observing relaxation from an ordered state to an equilibrium state.
This method is especially suitable for systems with large fluctuation and long relaxation, 
\textit{e.g.}, frustrated and random systems~\cite{ozeki01}.

Other than Monte Carlo simulations, DMRG is also popular, having been very successful in one-dimensional quantum systems~\cite{schollwoeck05}. 
Recently, developments in the field of quantum information have stimulated
extensions of DMRG to higher-dimensional systems.
Tensor-network algorithms are such extension~\cite{orus14, schollwoeck11, verstraete08}. 
One of the reasons why tensor-network algorithms  have called attention is that they can handle systems 
with large degrees of freedom with small computational cost as long as the system is homogeneous. 
For example, static critical exponents of two- and three-dimensional Ising models have been obtained in high accuracy by tensor renormalization group methods~\cite{levin07, zhao10, xie12, xie09}.

DMRG studies of  stochastic processes~\cite{hieida98, carlon99, carlon01, nagy02, kemper01, temme10} have been limited to one-dimensional systems until recently.
T. H. Johnson \textit{et al.} studied nonequilibrium stochastic processes in one and two dimensions using a tensor-network algorithm called time-evolving block decimation~\cite{johnson10, johnson15, vidal04}.
It discretizes the time of the dynamics of a finite system, using  the Suzuki-Trotter decomposition~\cite{suzuki76}.
They showed that their algorithm can compute in high accuracy large-variance observables that strongly depend on the time-evolving path of configuration, while Monte Carlo methods require a large number of samples for such variables.

The data structure of tensor networks represents the  spatial structure of the system and its correlation.
It is thus suited to computation of time evolution of systems with spatial correlations. 
Dynamical critical phenomena are examples of such systems. 
The dynamical critical exponent is a quantity that characterizes dynamical critical phenomena.
We are usually interested in dynamical critical phenomena for systems with dimensions greater than two 
because the dynamical critical exponents take nontrivial values there.
Their analytical calculation is usually intractable, 
and we need to rely on numerical methods.

% The data structure of tensor networks can naturally implement the information of correlation as well as it reflects the geometry of a lattice in a higher dimension.
% Discrete-time Markov chains often appear in statistical mechanics.
% For example, Monte Carlo simulations themselves are important examples of such processes. 
% Dynamical critical exponents are defined for them in the thermodynamic limit and we are often interested in systems greater than two. 
% We thus need a tool to investigate dynamical properties of critical phenomena.
% To obtain dynamical critical exponents precisely with small computational cost, 
% developing tensor-network algorithms for discrete-time Markov chains in infinite systems is important.

In the present paper, we propose a tensor-network algorithm for discrete-time Markov chains in $d$-dimensional infinite homogeneous systems.
Representing the  probability distribution with a tensor-network state and the transition probability with a tensor-network operator, we map $d$-dimensional nonequilibrium processes to $(d+1)$-dimensional infinite tensor networks.
While other tensor-network algorithms for dynamics usually make use of the Suzuki-Trotter decomposition~\cite{johnson10, johnson15, vidal04}, 
we construct a tensor-network operator of the transition probability in an entirely different way. 
We construct a tensor-network operator for a sublattice-flip update, using a higher-order singular-value decomposition~\cite{lathauwer00}.
Taking advantage of the homogeneity of the systems, we treat infinite systems directly just as the infinite time-evolving block decimation algorithm~\cite{vidal07, orus08}.
The correlation length diverges in a critical system, 
for which we need to study a large system. 
Our algorithm is especially suitable for the study of dynamical critical phenomena 
because the computational complexity of our algorithm does not depend on the system size.

We analyze nonequilibrium relaxation of the magnetization of the one- and two-dimensional Ising models as an application of our algorithm.
In particular, we determine the dynamical critical exponent $z$ of the two-dimensional Ising model. 
Our algorithm of time evolution particularly goes well with the nonequilibrium-relaxation method, for which one prepares a large system and compute time evolution for a relatively short time.
Our method has common features; it also prepares a systems so large that it can be treated as the thermodynamic limit and computes the time evolution for a short period.

This paper is organized as follows. 
In Sec.~\ref{sec:modelAndNotation}, we explain notation and the model that we use as an example of the algorithm that we propose.
We consider the Glauber dynamics~\cite{glauber63} in discrete time throughout this paper.
We borrow notations from quantum mechanics, expressing the probability distribution as a ket and the transition probability as an operator. 
In Sec.~\ref{sec:oneDKineticIsing}, we explain how to construct a tensor-network operator for the transition probability, using a higher-order singular-value decomposition. 
We calculate the relaxation of the magnetization from the all-spin-up state to the equilibrium with various bond dimensions and compare the results with an analytic result.
In Sec.~\ref{sec:twoDKineticIsing}, we analyze a system at the critical point and estimate the dynamical critical exponent $z$ using the nonequilibrium-relaxation method.

\section{Model and Notation \label{sec:modelAndNotation}}
Throughout this paper, we focus on kinetic Ising models, whose definition we give in this section;
we can easily generalize our algorithm to any sublattice-update dynamics with nearest-neighbor interactions on a bipartite graph.
We first define the update rule of spins and next explain how to describe a Markov process, borrowing the notation of quantum mechanics and using diagrams of tensor networks.

Unlike continuous time evolution, we do not need to perform the Suzuki-Trotter decomposition for the time evolution; we instead construct an operator of time evolution using the higher-order singular-value decomposition.   
  
\subsection{Kinetic Ising model \label{sec:glauber}}
The Glauber dynamics~\cite{glauber63} is a kind of kinetic Ising model whose transition rule is a heat-bath type. 
We consider the Glauber dynamics in discrete time although the original Glauber dynamics is in continuous time.
The Glauber dynamics in a computer simulation is usually the one in discrete time because such dynamics is simulated by Monte Carlo methods in many cases.
Our strategy is to implement heat-bath algorithms using tensor networks. 

We explain the Glauber dynamics taking the one-dimensional case as an example; generalizing it to higher dimensions is straightforward.
Let us consider a one-dimensional spin chain of length $2N$ whose Hamiltonian is given by 
\begin{align}
H = -\sum_{i=1,\cdots, 2N} \sigma_i \sigma_{i+1}.
\end{align}
We use periodic boundary conditions throughout this paper.

We update a single spin with the following heat-bath-type transition probability:
\begin{align}
w(\sigma_i \to \sigma_{i}') = \frac{e^{ \beta  \sigma_{i}'(\sigma_{i-1} + \sigma_{i+1}) }} { \sum_{\sigma_{i}''=\pm 1} e^{ \beta \sigma_{i}''(\sigma_{i-1} + \sigma_{i+1}) }}. \label{eqn:trSingleProb}
\end{align}
Here, $\sigma$ denotes a spin variable, which takes the values ${\pm 1}$,   
$i$ is the spin that we try to update, and $\sigma_i$ and $\sigma_{i}'$ are the values of the spin $i$ at the previous and new time steps, respectively. 
We adopt the two-sublattice multi-spin-flip dynamics, in which we divide the whole system into two sublattices and flip all the spins on each sublattice simultaneously.
In the odd time steps, the transition probability of the whole system is given as follows:
\begin{align}
w(\sigma_1,\cdots, \sigma_{2N}; \sigma_{1}', \cdots, \sigma_{2N}') 
= \prod_{i=1,3,\cdots, 2N-1} \frac{e^{ \beta  \sigma_{i}'(\sigma_{i-1} + \sigma_{i+1}) }} { \sum_{\sigma_{i}''=\pm 1} e^{ \beta \sigma_{i}''(\sigma_{i-1} + \sigma_{i+1}) }} \prod_{i=2,4,\cdots,2N}\delta_{\sigma_i, \sigma_{i}'} , \label{eqn:trOpOdd1D}
\end{align}
where $\delta_{ij}$ is Kronecker's delta.
The above transition probability means that spins at even sites are frozen and act as a heat bath to odd spins during the time evolution. 
This is a sufficient condition for relaxation to the equilibrium.
In the even time steps, roles of spins at odd and even sites are exchanged. 
The transition probability becomes  
\begin{align}
\label{eqn:trOpEven1D}
w(\sigma_1,\cdots, \sigma_{2N}; \sigma_{1}', \cdots, \sigma_{2N}') 
= \prod_{i=2,4,\cdots, 2N} \frac{e^{ \beta  \sigma_{i}'(\sigma_{i-1} + \sigma_{i+1}) }} { \sum_{\sigma_{i}''=\pm 1} e^{ \beta \sigma_{i}''(\sigma_{i-1} + \sigma_{i+1}) }} \prod_{i=1,3,\cdots,2N-1}\delta_{\sigma_i, \sigma_{i}'}. 
\end{align}
We obtain the probability distribution by applying the transition operators \eqref{eqn:trOpOdd1D} and \eqref{eqn:trOpEven1D} alternatively.

\subsection{Diagramattic representation of the Markov chain}
\begin{figure}
\begin{center}
\includegraphics[width=0.9\columnwidth]{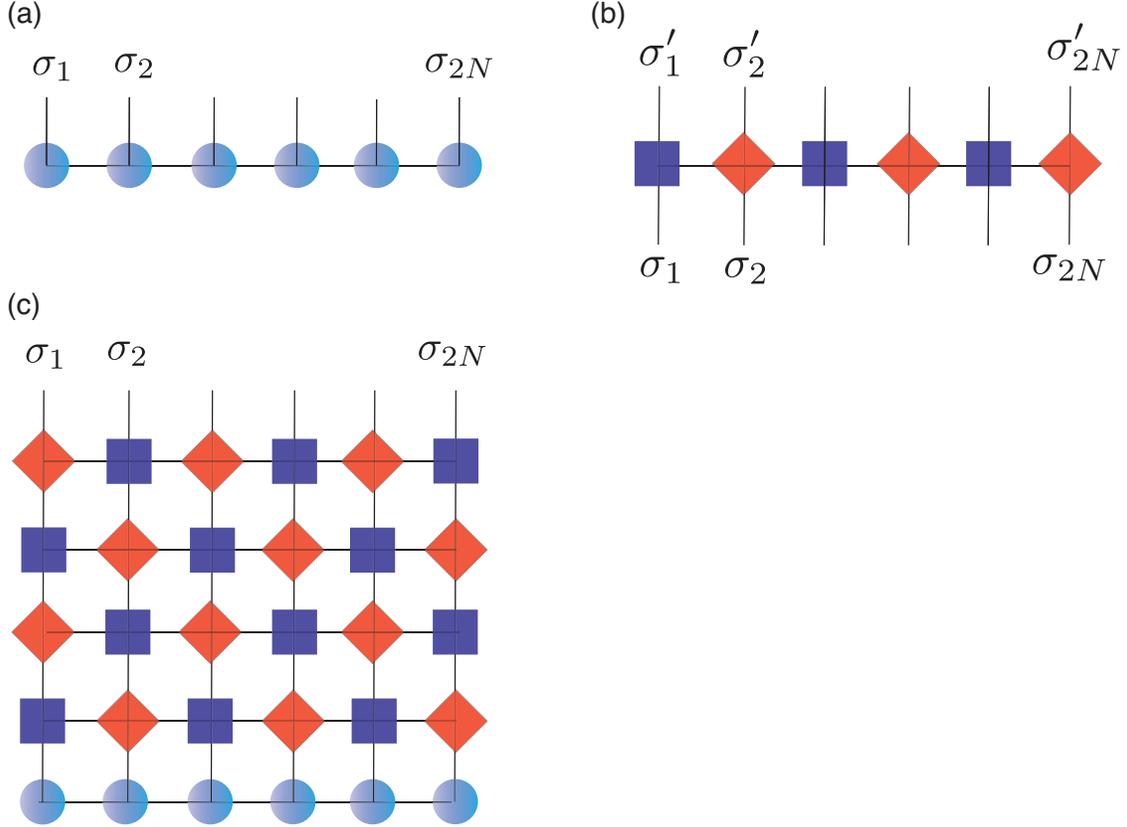}
\caption{(Color online) Diagrams in the tensor-network algorithm for a one-dimensional system. 
(a) A matrix-product-state representation of the probability distribution. (b) A matrix-product-operator representation of the transition probability. The squares and rhombuses are $Y$ and $X$ tensors defined in Eqs.~\eqref{eq:YopDef} and \eqref{eq:XopDef}, respectively.  (c) A diagrammatic representation of the time evolution~\eqref{eqn:timeEvolutionT}. 
\label{fig:1dFigs}}
\end{center}
\end{figure}

Consider a spin chain of length $M~~(M\in \mathbb{N})$ and denote the spin variables as $\vec{\sigma} = (\sigma_1, \cdots, \sigma_M)$. We express the probability distribution $P(\vec{\sigma})$ of a spin configuration borrowing the notation of quantum mechanics:
\begin{align}
P(\vec{\sigma}) = \langle \vec{\sigma} | P\rangle .
\end{align}
We now describe the ket of the probability distribution using a matrix-product state~\cite{schollwoeck11}:
\begin{align}
| P\rangle 
= \sum_{\vec{\sigma}} \mathrm{Tr} \left[ A^{[1]\sigma_1} \cdots A^{[M]\sigma_M}\right] |\sigma_1\cdots\sigma_M  \rangle . \label{eq:defOfMPS}
\end{align}
We can represent this matrix-product state as a diagram in Fig.~\ref{fig:1dFigs} (a).
For example, consider the fully magnetized state, in which all spins point up. 
We can represent it as a product of matrices of $1\times 1$:
\begin{align}
A^{[i]\sigma_i=+1} = 1,~~~A^{[i]\sigma_i=-1} = 0~~~(i=1,\cdots,M). \label{eq:allupMPS}
\end{align}
For another instance, let us consider a fully magnetized state whose spins are all up or down with the same probability. 
In this case, the state is more complex than the previous example, and we cannot express it as a product of $1\times 1$ matrices; 
we need $2\times 2$ matrices to express the state as a matrix product:
\begin{align}
A^{[i]\sigma_i=+1} = 2^{-1/M}\begin{pmatrix}1 & 0 \\ 0 & 0\end{pmatrix} ,
~~~A^{[i]\sigma_i=-1} = 2^{-1/M}\index{}\begin{pmatrix}0 & 0 \\ 0 & 1\end{pmatrix}~~~(i=1,\cdots,M)
\end{align}
Generally speaking, the more complex a state becomes, the larger matrices we need to express it as a matrix-product state.
The dimensionality of matrices to express a state is called the bond dimension of a matrix-product state; 
the bond dimension of the first example is thus one and that of the second example is two.

Next, consider the transition probability $W(\vec{\sigma}\to \vec{\sigma'})$ from a configuration $\vec{\sigma}$ to a new configuration $\vec{\sigma}'$;
it is a rank-$2M$ tensor, which we can represent as a matrix-product operator:
\begin{align}
\hat{W} 
= \sum_{\vec{\sigma}, \vec{\sigma}'} |\vec{\sigma}'\rangle \langle \vec{\sigma}'| \hat{W} | \vec{\sigma}\rangle  \langle \vec{\sigma}| :=\sum_{\vec{\sigma}, \vec{\sigma}'} W(\vec{\sigma}\to\vec{\sigma}') |\vec{\sigma}'\rangle \langle \vec{\sigma}| .
\end{align}
Figure~\ref{fig:1dFigs} (b) is a diagrammatic representation of this matrix-product operator.
Combining it with the matrix-product-state representation of the probability distribution, 
we can represent the time evolution of a probability distribution in the form 
\begin{align}
|P(t+1)\rangle = \hat{W} | P(t)\rangle \label{eqn:timeEvolutionSingle}, 
\end{align}
which is followed by 
\begin{align}
|P(t)\rangle = \hat{W}^t | P(0)\rangle. \label{eqn:timeEvolutionT}
\end{align}
The corresponding diagram is Fig.~\ref{fig:1dFigs} (c).
%When dimensions of every connected bond is less than $D$, then the tensor network is called to have bond dimensions $D$.

\begin{figure}
\begin{center}
\includegraphics[width=0.9\columnwidth]{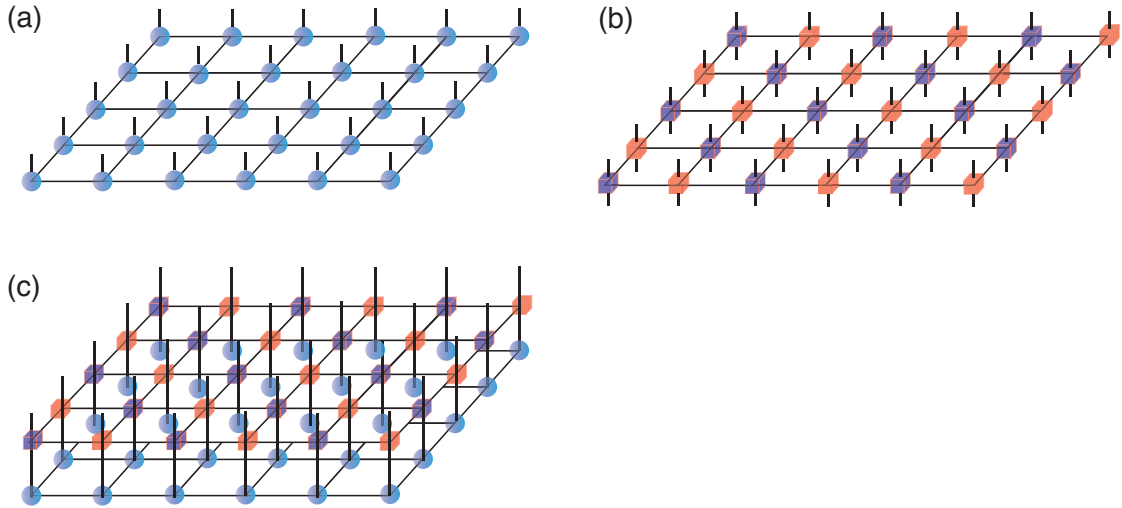}
\caption{(Color online) Diagrams in the tensor-network algorithm for a two-dimensional system. 
(a) A tensor-network-state representation of the probability distribution. (b) A tensor-network-operator representation of the transition probability. (c) Time evolution in a single time step, which updates half of the spins (Eq.~\eqref{eqn:timeEvolutionSingle}).
\label{fig:2dFigs}}
\end{center}
\end{figure}

In a two-dimensional system, the probability distribution and the transition probability become a tensor-network state and a tensor-network operator, respectively (Fig.~\ref{fig:2dFigs}).
%As is well known, $d$-dimensional nonequilibrium system is equivalent to another $(d+1)$-dimensional system and this correspondence is now established by using tensor networks. f
We can represent any transition probabilities with tensor-network operators in principle. 
However, writing down its form is not a trivial problem in practice and requires ingenuity. 
For the two-sublattice multi-spin-flip dynamics, we can write down the tensor-network operator explicitly as we will explain in the following two sections.

\section{One-dimensional kinetic Ising model \label{sec:oneDKineticIsing}}
We numerically analyze the one-dimensional kinetic Ising model in this section.
The update rule is the one explained in Sec.~\ref{sec:glauber}. 
We prepare all the spins to be $\sigma_i = 1$ at the initial time and observe the relaxation to equilibrium. 
N. Ito \textit{et al.} derived an asymptotic form of the relaxation of the magnetization analytically~\cite{ito90a,ito90b}.
We calculate it numerically and compare the result with their asymptotic form.

\subsection{Transition matrix as matrix-product operator}
We use the transition probabilities given by the rank-$4N$ tensors~\eqref{eqn:trOpOdd1D} and \eqref{eqn:trOpEven1D}.
We first decompose the local transition probability by using the higher-order singular-value decomposition~\cite{lathauwer00, kolda09} as 
\begin{align}
w(\sigma_i \to \sigma_{i}') 
=\frac{e^{ 1/T  \times \sigma_{i}'(\sigma_{i-1} + \sigma_{i+1}) }} { \sum_{\sigma_{i}''=\pm 1} e^{ 1/T \times \sigma_{i}''(\sigma_{i-1} + \sigma_{i+1}) }}
= \sum_{\alpha,\beta,\gamma=1,2} S_{\alpha\beta\gamma} U_{\sigma_{i}'\alpha} V^{(L)}_{\sigma_{i-1}\beta} V^{(R)}_{\sigma_{i+1}\gamma},
\label{eqn:hosvd1d}
\end{align}
\begin{figure}
\begin{center}
\includegraphics[width=0.3\columnwidth]{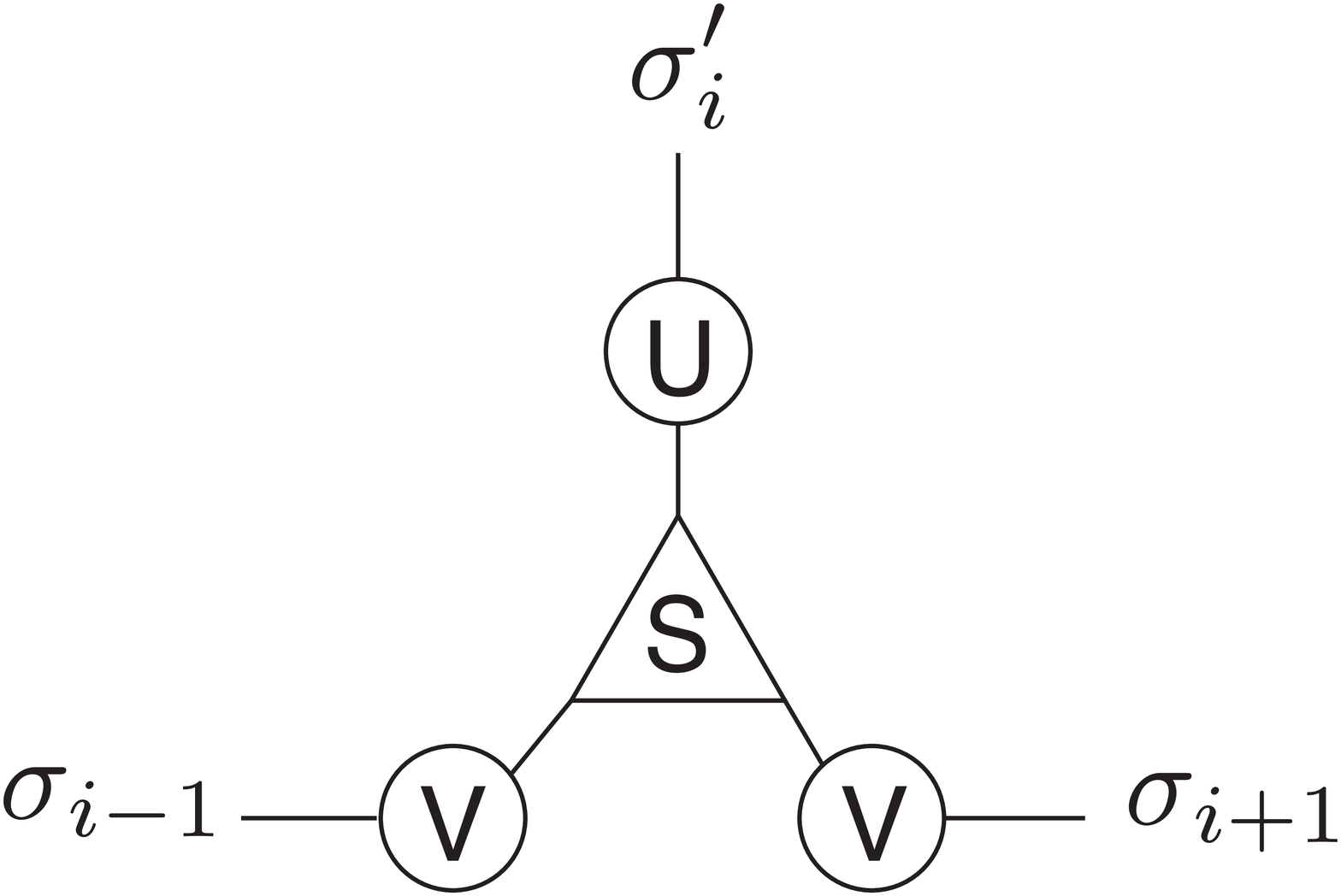}
\caption{A diagrammatic representation of the higher-order singular-value decomposition (HOSVD) of a local transition probability Eq.~\eqref{eqn:hosvd1d}.
\label{fig:hosvd1d}}
\end{center}
\end{figure}
where $S$ is a rank-3 tensor called the core tensor, while $U$, $V^{(R)}$, and $V^{(L)}$ are $2\times 2$ orthogonal matrices. 
We performed this decomposition numerically.
We can represent this equation using a diagram in Fig.~\ref{fig:hosvd1d}.
We next define the following rank-4 tensors:
\begin{align}
X^{\sigma_{j}'\sigma_j}_{pq} &:= V^{(R)}_{\sigma_{j} p} V^{(L)}_{\sigma_{j}q}\delta_{\sigma_{j}' \sigma_j} ,\label{eq:XopDef}\\
Y^{\sigma_{i}'}_{\beta \gamma} &= Y^{\sigma_{i}'\sigma_i}_{\beta \gamma} := \sum_{\alpha=1,2}S_{\alpha\beta\gamma}U_{\sigma_{i}'\alpha}  . \label{eq:YopDef}
\end{align}
We let $Y$ have an index $\sigma_i$ because we need $Y$ to have both indices $\sigma_i$ and $\sigma_{i}'$ in the definition of a matrix-product operator below. 
We obtain matrix-product-operator representations of the whole system by lining up $X$ and $Y$ (Fig.~\ref{fig:1dFigs} (b)).
For instance, in the odd time steps,
\begin{align}
W(\vec{\sigma}\to\vec{\sigma'}) = \mathrm{Tr}[Y^{\sigma_{1}'\sigma_{1}} X^{\sigma_{2}'\sigma_{2}} Y^{\sigma_{3}'\sigma_{3}} X^{\sigma_{4}'\sigma_{4}} \cdots].
\end{align}
In the even time steps, the order of $X$ and $Y$ in the trace is interchanged.
We thus represent the time evolution by piling up the two matrix-product operators alternatively to the initial matrix-product state (Fig.~\ref{fig:1dFigs} (c)).
We set the initial state to be the all-spin-up state, which is a matrix-product state of bond dimension $D=1$. 
The bond dimensionality of the matrix-product operator is two.

We calculate $|P(t+1)\rangle$ from $|P(t)\rangle$ using Eq.~\eqref{eqn:timeEvolutionSingle}.
We can represent $|P(t+1)\rangle$ as a matrix-product state since $|P(t)\rangle$ is also a matrix-product state and $\hat{W}$ is a matrix-product operator.
Our initial state $|P(0)\rangle$ is the matrix-product state of Eq.~\eqref{eq:allupMPS}.
We can represent the procedure of  time evolution graphically.
Suppose  that Fig.~\ref{fig:1dFigs} (a) is the state at $t=0$. 
Applying the matrix-product operator of Fig.~\ref{fig:1dFigs} (b), we obtain the state at $t=1$. 
Repeating the procedure, we can temporally evolve the stochastic process.
For example, Fig.~\ref{fig:1dFigs} (c) is the state at $t=4$. 
In numerical computation, we contract vertical bonds every time we apply a matrix-product operator to a state. 
Therefore, states are always expressed as matrix-product states. 
The shape of the state is the same as in Fig.~\ref{fig:1dFigs} (a) but all tensors are not the same for $t\ge 1$  because the applied matrix-product operators consist of the two kinds of matrices~\eqref{eq:XopDef} and ~\eqref{eq:YopDef}. 
Each time we apply a matrix-product operator to a state, the bond dimensionality of the state doubles, as the bond dimensionality of the matrix-product operator is two. 
We could be able to express the state at $t$ exactly as a product of matrices of size $2^t$.
However, as the amount of memory is limited, of course, 
we need to perform approximation;
we truncate the bond dimensions  at a dimensionality $D$ by using the singular-value decompositions and 
keep the $D$ largest singular values. 
When the dimensionality of every connected bond is less than $D$, we say that the tensor network has bond dimensions $D$.

We explain this procedure in detail here, following Ref.~\cite{schollwoeck11}. 
Let us consider approximating the bond connected by the two  tensors $A^{[i]}$ and  $A^{[i+1]}$ in Eq.~\eqref{eq:defOfMPS}.
The product before approximation is 
\begin{align}
\sum_{j=1}^{D'} A_{i,j}^{[i]\sigma_s} A_{j,k}^{[i+1]\sigma_t}. \label{eq:STproduct}
\end{align}
The  dimensions of the indices $\sigma_s$ and $\sigma_t$ are two, and we assume that  the dimensions of the indices $i,j,$ and $k$ are $D'$ ($D' > D$). 
Our task is to approximate the contraction~\eqref{eq:STproduct}, which is the summation of $D'$ terms, by a summation of $D$ terms. 
First, we rewrite the tensors with three indices to  tensors with two indices (matrices):
\begin{align}
A_{i,j}^{[i]\sigma_s}=A^{'[i]}_{(i,\sigma_s), j},~~~A_{j,k}^{[i+1]\sigma_t}=A^{'[i+1]}_{j, (k,\sigma_t)}.
\end{align}
Then the contraction \eqref{eq:STproduct} becomes a multiplication of the new two matrices: $A^{'[i]} A^{'[i+1]}$. 
We now approximate the product of two rank-$D'$ matrices by a product of two rank-$D$ matrices as follows. 
We first decompose $A^{'[i]} A^{'[i+1]}$ by using the singular-value decomposition:
\begin{align}
A^{'[i]} A^{'[i+1]}=U\Lambda V^{\dagger} .
\end{align}
Here $\Lambda$ is a diagonal matrix of size $D'$ and we assume that the eigenvalues are sorted in the non-ascending order. 
We keep only the $D$ largest eigenvalues of $\Lambda$ and omit the other small eigenvalues: 
\begin{align}
U\Lambda V^{\dagger} \approx \tilde{U} \tilde{\Lambda} \tilde{V}^{\dagger},
\end{align}
where $\tilde{\Lambda}$ is the diagonal matrix of size $D$. 
We left out the last $D'-D$ columns of $U$ and $V$, and defined the two new matrices $\tilde{U}$ and $\tilde{V}$. 
We now approximate $A^{'[i]}$ and $A^{'[i+1]}$ as follows:
\begin{align}
A^{'[i]}  & \approx \tilde{A}^{'[i]} :=  \tilde{U}\sqrt{\tilde{\Lambda}},\\
A^{'[i+1]}  & \approx \tilde{A}^{'[i+1]} := \sqrt{\tilde{\Lambda}}\tilde{V}^{\dagger}.
\end{align}
We reshape the matrices $\tilde{A}^{'[i]}_{(i,\sigma_s),j}$ and $\tilde{A}^{'[i+1]}_{j,(k,\sigma_t)}$ to tensors with three indices again. 
\begin{align}
\tilde{A}^{'[i]}_{(i,\sigma_s),j}& = \tilde{A}_{i,j}^{'[i]\sigma_s},\\
\tilde{A}^{'[i+1]}_{j,(k,\sigma_t)}& = \tilde{A}_{j,k}^{'[i+1]\sigma_t} ,
\end{align}
where $i$ and $k$ runs over $1,\cdots, D'$, $j$ runs over $1,\cdots,D$, and $\sigma_s$ and $\sigma_t$ takes $\pm 1$. 
The contraction of the $D'$ terms are finally approximated by the contraction of the $D$ terms:
\begin{align}
\sum_{j=1}^{D'} A_{i,j}^{[i]\sigma_s} A_{j,k}^{[i+1]\sigma_t}\approx 
\sum_{j=1}^{D} \tilde{A}_{i,j}^{[i]\sigma_s} \tilde{A}_{j,k}^{[i+1]\sigma_t}.
\end{align}
The computational complexity of the time evolution is of $O(D^3)$
because the singular-value decomposition takes CPU time of  $O(m n^{2})$ for an $m\times n~(m\ge n)$ matrix.

\begin{figure}
\begin{center}
\includegraphics[width=0.7\columnwidth]{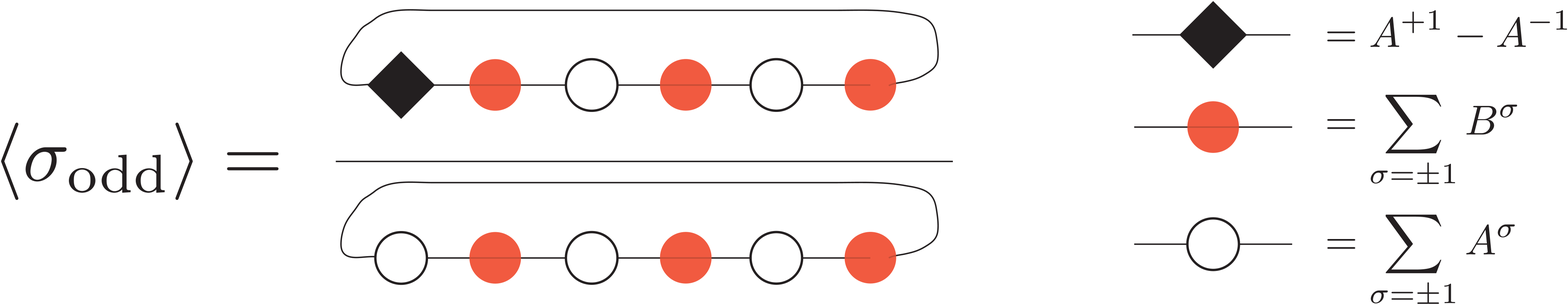}
\caption{(Color online) A diagrammatic representation of the average of the spins at odd sites.
The denominator is the norm of probability. 
The right figure shows the definition of each symbol.
\label{fig:calcMean1d}}
\end{center}
\end{figure}

We finally calculate the average magnetization at the odd sites and that at the even sites separately.
Let us consider calculating the average of $\sigma_1$ as a representative of the odd spins.
Suppose that the state at time $t$ is $|P(t)\rangle=\sum_{\vec{\sigma}}\mathrm{Tr}[A^{\sigma_1}B^{\sigma_2}A^{\sigma_3}B^{\sigma_4}\cdots ]|\vec{\sigma}\rangle$.
%To calculate the average magnetization at odd sites, 
We first compute the marginal distribution of $|\sigma_1\rangle$:
 %we calculate the mean of the first spin.
%We can calculate the average magnetization at odd sites as follows.
\begin{align}
P(\sigma_1; t) := \sum_{\sigma_2,\cdots,\sigma_{2N}}\mathrm{Tr}[A^{\sigma_1}B^{\sigma_2}A^{\sigma_3}B^{\sigma_4}\cdots ].
\end{align}
The average magnetization of $\sigma_1$ is then given by
\begin{align}
\langle \sigma_1\rangle = \sum_{\sigma_1=\pm 1} \sigma_1 P(\sigma_1;t) 
= \mathrm{Tr}[(A^{+1}-A^{-1}) \sum_{\sigma_2}B^{\sigma_2}  \sum_{\sigma_3}A^{\sigma_3}  \sum_{\sigma_4}B^{\sigma_4}\cdots ] .
\end{align}
We can represent this equation as the diagram shown in Fig.~\ref{fig:calcMean1d}.
Each open circle denotes $\sum_{\sigma=\pm 1}A^{\sigma}$, each solid circle denotes $\sum_{\sigma=\pm 1}B^{\sigma}$, and the solid rhombus denotes $A^{+1} - A^{-1}$. 

We repeat a sufficient number of renormalization until the result converges.
We explain the procedure of renormalization in detail.
In the first step, we define 
\begin{align}
C&=(A^{+1} - A^{-1})\sum_{\sigma=\pm 1}B^{\sigma} , \\
E& =\sum_{\sigma_a=\pm 1}A^{\sigma_a}\sum_{\sigma_b=\pm 1}B^{\sigma_b}.
\end{align} 
In the second step, namely the renormalization step, we update $C$ and $E$ as \begin{align}
C&\leftarrow CE,\\
E&\leftarrow EE.
\end{align}
The sum of probabilities $\sum_{\vec{\sigma}}\mathrm{Tr}[A^{\sigma_1}B^{\sigma_2}A^{\sigma_3}B^{\sigma_4}\cdots ]$ deviates from unity as time evolves because of truncations by singular-value decompositions.
To preserve the normalization of the probability, we divided the magnetization by the norm of the probability   
distribution at time $t$ as in Fig.~\ref{fig:calcMean1d}.
The average magnetization at odd sites is 
\begin{align}
\langle \sigma_{\mathrm{odd}} \rangle = \frac{\operatorname{Tr}C}{\operatorname{Tr}E}.
\end{align}
We repeat this renormalization procedure until the average magnetization $\langle \sigma_{\mathrm{odd}} \rangle $ converges.

The calculation of the magnetization at  even sites is similar.
The magnetization of the whole system is the average of these two averaged magnetizations.
The evaluation of the diagram in Fig.~\ref{fig:calcMean1d} takes CPU time of $O(D^3)$.
The total computational complexity of our algorithm is thus of $O(D^3)$.

\begin{figure}
\begin{center}
\includegraphics[width=0.7\columnwidth]{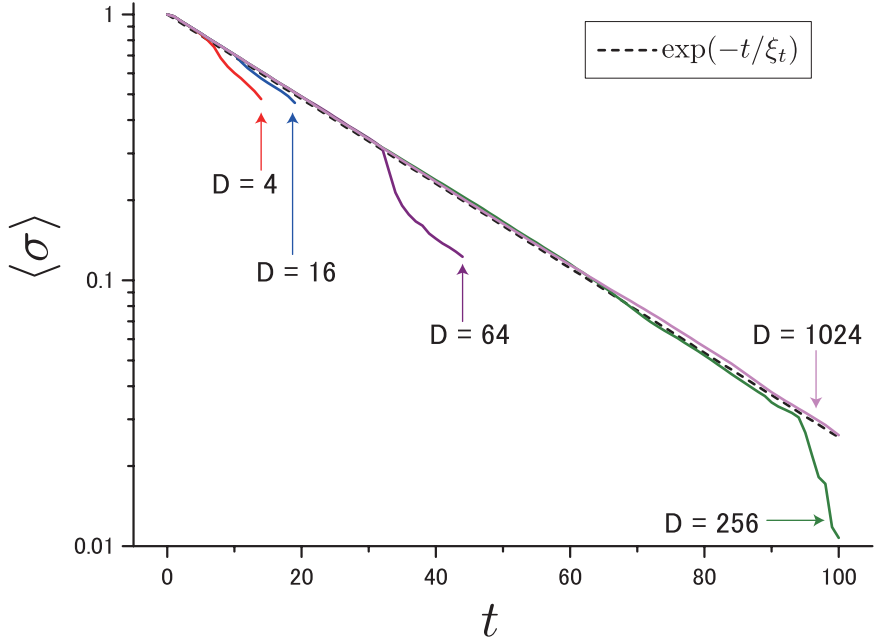}
\caption{(Color online) The relaxation of the magnetization of the one-dimensional Ising model. 
The horizontal axis indicates the number of the time steps, while the vertical axis indicates the magnetization.
We changed the maximum bond dimension $D$. 
The broken line is the analytic asymptotic form~\cite{ozeki07};
the magnetization decays exponentially with the correlation time $\xi_t$ defined in Eq.~\eqref{eq:xi_t1d}. 
\label{fig:oneDIsingMag}}
\end{center}
\end{figure}

We show the results of the calculation in Fig.~\ref{fig:oneDIsingMag}. 
The broken line in the figure is the analytic asymptotic form~\cite{ito90a,ito90b};
the magnetization decays exponentially with the correlation time
\begin{align}
\xi_t = \frac{1}{\log(\coth(2/T))}. \label{eq:xi_t1d}
\end{align} 
As the bond dimension $D$ increases, the range of time for which we can calculate the magnetization precisely also expands. 
We were able to do so up to around $t\approx 100$ in one dimension with $D=1024$.

We update half of the tensors at each step of time. 
%This corresponds with a 1/2 Monte Carlo step in the case of Monte Carlo simulations. 
An advantage of our algorithm is that we can update half of the system at the same time, making use of the translational variance of the system.
We just need to update a single tensor of a sublattice because the tensors of the same sublattice are all the same. 
In Monte Carlo simulation, on the other hand,  
flipping a half of the system takes the CPU time that increases linearly in the system size.
\section{Two-dimensional kinetic Ising model \label{sec:twoDKineticIsing}}
We numerically analyze the two-dimensional kinetic Ising model in this section.
In particular, by using the nonequilibrium-relaxation method, we estimate the dynamical critical exponent $z$ of the two-dimensional Ising model.

\subsection{Transition operator as tensor-network operator}
The update rule of a single spin in the one-dimensional Glauber dynamics, Eq.~\eqref{eqn:trSingleProb}, changes in the two-dimensional case to
\begin{align}
w(\sigma_i \to \sigma_{i}') &= \frac{e^{ 1/T \times \sigma_{i}'(\sigma_{L} + \sigma_{R} + \sigma_{D} + \sigma_{U}   ) }} { \sum_{\sigma_{i}''=\pm 1} e^{ 1/T \times \sigma_{i}''(\sigma_{L} + \sigma_{R} + \sigma_{D} + \sigma_{U} ) }}  \\
&= \sum_{\alpha,\beta,\gamma,\delta,\epsilon=1,2}   S_{\alpha\beta\gamma\delta\epsilon} U_{\sigma_{i}' \alpha} V^{(L)}_{\sigma_L \beta}  V^{(R)}_{\sigma_R \gamma}  V^{(D)}_{\sigma_D \delta}  V^{(U)}_{\sigma_U \epsilon},
\end{align}
where the subscripts $L, R, U,$ and $D$ represent spins to the left, right, up, and down of the spin $\sigma_i$, respectively. 
As in the one-dimensional case, we perform the higher-order singular-value decomposition (HOSVD) and 
define local transition operators:
\begin{align}
X_{pqrs}^{\sigma_{j}'\sigma_j} &:= V^{(R)}_{\sigma_j p} V^{(L)}_{\sigma_j q} V^{(U)}_{\sigma_j r} V^{(D)}_{\sigma_j s} \delta_{\sigma_{j}' ,\sigma_j},\\
Y_{\beta\gamma\delta\epsilon}^{\sigma_{i}'} 
&:= Y_{\beta\gamma\delta\epsilon}^{\sigma_{i}' \sigma_i}
= \sum_{\alpha=1,2} S_{\alpha\beta\gamma\delta\epsilon} U_{\sigma_{i}' \alpha} .
\end{align}
A diagrammatic representation of the transition operator of the whole system consists of the local transition operators $X$ and $Y$.
The bond dimensionality of the transition operator is two;
see Fig.~\ref{fig:2dFigs} (b).
We calculate the time evolution by stacking up the tensor-network operators for each of odd and even time steps alternatively (Fig.~\ref{fig:2dFigs} (c)).

We calculate the time evolution of a state by contracting the state and tensor-network operators from the bottom layers.
The state at $t=0$ is shown in Fig.~\ref{fig:2dFigs} (a). 
Applying the tensor-network operator of Fig.~\ref{fig:2dFigs} (b) to this state, we can evolve the dynamics by one step (Fig.~\ref{fig:2dFigs} (c)). 
We then contract vertical bonds of Fig.~\ref{fig:2dFigs} (c) and obtain the tensor-network state at $t=1$.
The shape of the tensor network state at $t=1$ is the same as  Fig.~\ref{fig:2dFigs} (a), but all tensors are not the same.
The lattice consists of two sublattices; 
tensors of each sublattice share the common tensor.  
By repeating the same procedure, we can calculate the states of $t \ge 1$, $|P(t)\rangle$.
During the calculation of the time evolution,
we truncate bond indices by singular-value decompositions if the bond dimensionality exceeds $D$.  
We can do these singular-value decompositions in the time of $O(D^{5})$ with a little ingenuity~\cite{wang14}. 

We finally compute the magnetization by repeating renormalization of the tensor network that contains an ``impurity tensor" (the solid rhombus in Fig.~\ref{fig:calcMean1d}) until convergence as in the one-dimensional case.
We also divide the magnetization, which corresponds to the numerator of Fig.~\ref{fig:calcMean1d}, by the norm of $|P(t)\rangle$ (the denominator of the same figure) as in the one-dimensional case.
We use the algorithm by C. Wang \textit{et al.}~\cite{wang14}, 
in which one truncates bond dimensions by singular-value decompositions during renormalization, 
whereas the other parts of the procedure are the same as the tensor renormalization group with the higher-order singular-value decomposition~\cite{xie12}.
The calculation of the average magnetization by tensor renormalization group takes the time of $O(D^{8})$~\cite{wang14}
and the total computational complexity of our algorithm in two dimensions is also of $O(D^{8})$.

\subsection{Nonequilibrium relaxation}
We observe the relaxation from the all-spin-up initial state to the equilibrium state. 
The initial state is all-spin-up, which is a two-dimensional tensor-network state of the bond dimension $D=1$.
The asymptotic behavior of the magnetization depends on which phase the system is in.
In our case, since we do not apply a magnetic field, the asymptotic decay becomes as follows:
\begin{eqnarray}
\langle \sigma \rangle = 
\begin{cases}
e^{-t/\xi_t} ~~~&T>T_c, \\
t^{-\lambda_m} ~~~&T=T_c, \\
m_{\mathrm{eq}} + c e^{-t/\xi_t} ~~~&T<T_c,
\end{cases}
\end{eqnarray}
where $T_c=2.269$, $\xi_t$ is the relaxation time, $m_{\mathrm{eq}}$ is the spontaneous magnetization, $c$ is a constant, and 
$\lambda_m$ is the dynamical critical exponent that characterizes the power-law decay of the magnetization at the critical point. It is related with the standard critical exponents as ~\cite{ito93,ozeki07}
\begin{align}
\lambda_m = \frac{\beta} { z \nu }.  
\end{align}
In the two-dimensional Ising model, the critical exponents $\beta$ and $\nu$ have been obtained analytically at $1/8$ and $1$, respectively, and therefore we can evaluate $z$ from the decay of the magnetization at the critical point.
This method is called the nonequilibrium-relaxation method~\cite{ito93,ozeki07}. 

\begin{figure}
\begin{center}
\includegraphics[width=1\columnwidth]{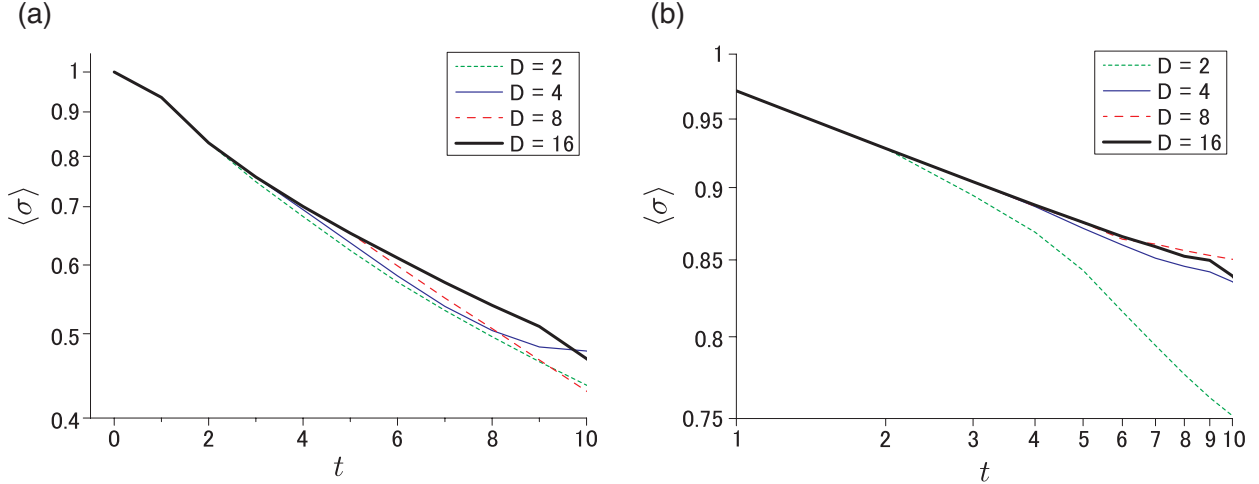}
\caption{(Color online) Relaxation of the magnetization from the all-spin-up state in the two-dimensional Ising model. 
The horizontal axis indicates the time steps and the $y$-axis indicates the  magnetization, while
$D$ denotes the bond dimensionality of tensors.
(a) A semi-logarithmic plot for $T=3>T_c$ under no magnetic field. 
%The $y$-axis is a log scale, and the $x$-axis is a linear scale.
(b) A logarithmic plot at the critical point $T=T_c\approx 2.269$. 
%Both $x$- and $y$-axises are log scales. 
\label{fig:relaxation2d}}
\end{center}
\end{figure}

\begin{figure}
\begin{center}
\includegraphics[width=0.7\columnwidth]{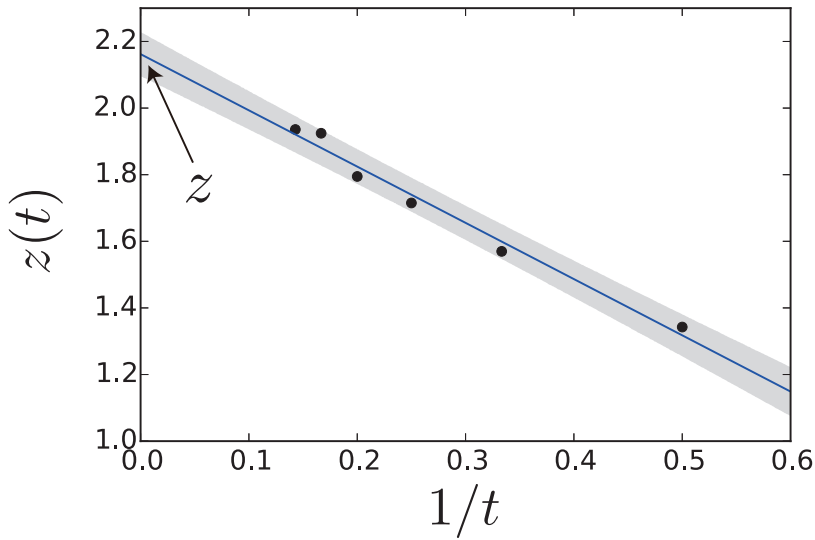}
\caption{Extrapolation of the local exponent $z(t)$ to $t\to\infty$ according to Eq.~\eqref{eqn:extrapolatezt}.
We performed the Bayesian linear regression.
The line of the best fit is the \textit{maximum a posteriori} solution and the shaded area is the 75 \% credible interval.
The intersection of the line of the best fit and the $y$-axis is the final estimate of the critical exponent $z$.
% We defined the uncertainty of $z$ as the 75 \% credible interval at the % origin of $x$ axis, which corresponds to $t\to\infty$.
The data is for the bond dimension of tensors $D=20$.
\label{fig:extrapolatezt}}
\end{center}
\end{figure}

We calculated the relaxation of systems in the high-temperature phase and at the critical point (Fig.~\ref{fig:relaxation2d}). 
In the high-temperature phase, the magnetization decays exponentially in time with the correlation time $\xi_t$, while
at the critical point $\xi_t$ diverges, and the magnetization shows a power-law decay. 
To calculate the critical exponent $z$ precisely, we calculated ``local exponents" and extrapolated the results to the limit of the infinite time~\cite{ito93, ozeki07}. 
We define the local exponents by
\begin{align}
\lambda_m(t) &:= -\frac{d\log m(t)} {d\log t} \approx \frac{t}{\Delta t} \left( \frac{m(t-\Delta t)}{m(t)} - 1\right),  \\
z(t) &:= \frac{\beta}{\nu \lambda_m(t)}. \label{eqn:defzt}
\end{align}
We choose $\Delta t = 1$ and fit the series $z(t)$ to 
\begin{align}
z(t) = a/t + z, \label{eqn:extrapolatezt}
\end{align}
where $a$ is a constant and $z$ is the final estimate of the critical exponent when the numbers of times steps are extrapolated to infinity.
We did the extrapolation by the Bayesian linear regression~\cite{gelman14} (Fig.~\ref{fig:extrapolatezt}).
The intersection of the line of the best fit and the $y$-axis is the point estimate of $z$.
% and we estimated the value to be $z=2.16$.
%The 75 \% credible interval at the origin of the $x$-axis was $[2.12,2.21]$,
We used the 75 \% credible interval at the origin of the $x$-axis as the uncertainty of our estimate of $z$.

\begin{figure}
\begin{center}
\includegraphics[width=0.7\columnwidth]{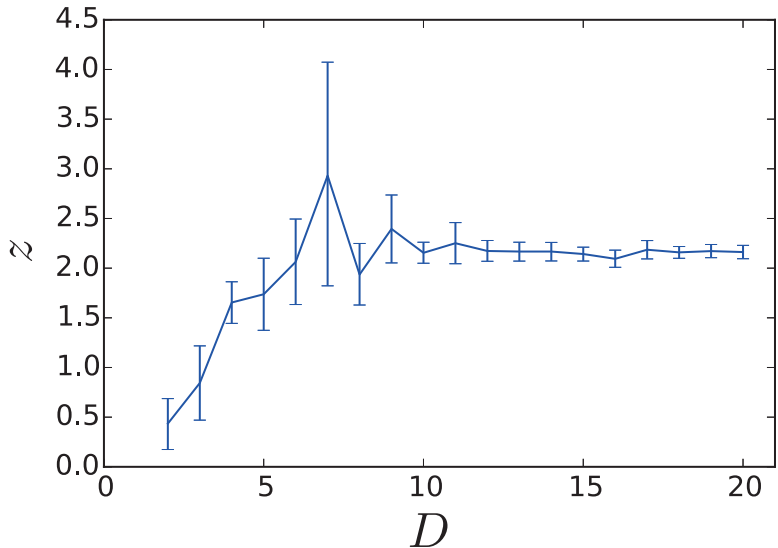}
\caption{Estimates of the dynamical critical exponent $z$ with various bond dimensions $D$. 
The error bars indicate the 75 \% credible intervals of  the predictive distribution  at $1/t=0$ in Eq.~\eqref{eqn:extrapolatezt}, $i.e.$, the shaded area at the origin of the $x$-axis in Fig.~\ref{fig:extrapolatezt}.
\label{fig:estimateOfz}  }
\end{center}
\end{figure}

We calculated $z$ for various bond dimensions ranging from $D=2$ to $D=20$ and carried out the analysis described as above. 
In the extrapolation of the local exponent $z(t)$, we did not use the first two data points corresponding to $t=0,1$.
We used five data points ($t=2,\cdots,6$) for $D<17$ and six data points ($t=2,\cdots,7$) for $D\ge 17$.
As Fig.~\ref{fig:estimateOfz} shows, the estimates of $z$ converges as $D$ increases. 
We estimated the critical exponents $z$ to be $2.16(5)$ using data of $D=20$.
The nonequilibrium-relaxation method with a Monte Carlo method~\cite{ito93} estimated $z$ to be $z = 2.165(10)$ and a series expansion~\cite{dammann95} did $z$ to be $z = 2.183(5)$. 
Our result is consistent with the values of these studies.

\section{Conclusion}
We have proposed an algorithm to compute sublattice-flip dynamics on an infinite bipartite graph. 
We split local transition probabilities into tensors that depend only on a single spin variable by a higher-order singular-value decomposition and constructed a tensor-network-operator representation of the transition probability of the whole system. 
We can apply this approach to any sublattice-update dynamics with nearest-neighbor interactions on a bipartite graph.
We treat an infinite system utilizing translational invariance and obtained the magnetization in the thermodynamic limit directly without system-size extrapolation.
Instead of extrapolating the system size to infinity, we did the bond-dimension extrapolation. 
We are also able to  apply our algorithm to an open boundary system in principle.
The system, however, becomes inhomogeneous, and hence we will need to keep all the tensors that lie on all the sites with the computational costs  depending on the system size. Therefore, studying dynamics in the thermodynamics limit directly is only possible  when we adopt the periodic boundary condition.

Our algorithm goes together well with the nonequilibrium-relaxation method.
In the nonequilibrium-relaxation method, one prepares a large system and simulate it for a rather small number of time steps~\cite{ozeki07}. 
In calculation of the time evolution with tensor networks, we cannot compute dynamics for an arbitrary long time because an error due to singular-value decompositions accumulates during the time evolution. 
We can, however, update an infinite number of spins at once utilizing translational invariance.
We calculated the time evolution of an infinite system for a short period indeed with good precision and were able to determine the critical exponent $z$.

We estimated only $z$ among critical exponents because the other exponents have been analytically known. 
The nonequilibrium-relaxation method, however, can calculate the other critical exponents as well as  the critical temperature, and thus we can use it even for systems for which analytical calculation is intractable. 
The nonequilibrium-relaxation method combined with our algorithm is a new direction of study of critical phenomena with tensor networks.

A shortcoming of our algorithm at present is that we can utilize it for a limited range of time. 
This problem is serious in two dimensions; we obtained the results only for $t<10$ because it became difficult to increase bond dimensions further as the dimensionality increased. 
The accuracy of our estimate of the critical exponent $z$ is worse than the estimate of the nonequilibrium-relaxation method with a Monte Carlo simulation~\cite{ito93} because we used a smaller number of time steps.
In order to compute longer and to improve the accuracy of the estimate, we need to develop a better scheme to truncate bond dimensions during the time evolution.
%Furthermore, there is a possibility to improve results using a recent tensor renormalization group algorithm, called the tensor network renormalization. 

\begin{acknowledgments}
I am grateful to my supervisor Prof. Naomichi Hatano for reading this manuscript. 
I also thank Prof. Tota Nakamura for introducing me to the nonequilibrium-relaxation method.
The work is supported by Advanced Leading Graduate Course for Photon Science.
\end{acknowledgments}

\bibliographystyle{apsrmp4-1} % use this to hide titles
\bibliography{2015MCbyTensorNet}

\end{document}